\documentclass{article}

\usepackage{PRIMEarxiv}

\usepackage[utf8]{inputenc} 
\usepackage[T1]{fontenc}    
\usepackage{hyperref}       
\usepackage{url}            
\usepackage{booktabs}       
\usepackage{amsfonts}       
\usepackage{nicefrac}       
\usepackage{microtype}      
\usepackage{lipsum}
\usepackage{fancyhdr}       
\usepackage{graphicx}       
\usepackage{xcolor}
\usepackage{listings}
\usepackage{geometry}
\usepackage[T1]{fontenc}  
\usepackage{lmodern}      
\usepackage{appendix}   
\usepackage{hyperref}   

\lstdefinestyle{plaintextstyle}{
    language={},
    basicstyle=\ttfamily\footnotesize,
    backgroundcolor=\color{gray!10},
    numbers=left,
    numberstyle=\tiny\color{gray},
    numbersep=5pt,
    breaklines=true,
    showstringspaces=false,
    frame=single,
    rulecolor=\color{black!30},
    framesep=5pt,
    xleftmargin=20pt,
    framexleftmargin=17pt,
    stepnumber=1,
    numbersep=5pt,
    firstnumber=1,
    numberblanklines=false,
    keywordstyle={},
    commentstyle={},
    stringstyle={}
}

\graphicspath{{media/}}     

\usepackage{appendix}

\title{MOSS: Enabling Code-Driven Evolution and Context Management for AI Agents}

\author{
  Ming Zhu\thanks{Ming Zhu previously worked at ByteDance's AI division, where he was one of the key designers of the multi-agent system for Coze, an AI chatbot development platform that supports multi-task handling.} \\
  Independent Researcher \\
  \texttt{thirdgerb@gmail.com}  \\
  \and
  Yi Zhou\thanks{First Author and Second Author contribute equally to this work.} \\
  Beihang University \\
  \texttt{nilezhou123@buaa.edu.cn} \\
}

\begin{document}
\maketitle

\begin{abstract}
Developing AI agents powered by large language models (LLMs) faces significant challenges in achieving true Turing completeness and adaptive, code-driven evolution. Current approaches often generate code independently of its runtime context, relying heavily on the LLM's memory, which results in inefficiencies and limits adaptability. Manual protocol development in sandbox environments further constrains the agent's autonomous adaptability. Crucially, achieving consistency in code and context across multi-turn interactions and ensuring isolation of local variables within each interaction remains an unsolved problem.   

We introduce MOSS (llM-oriented Operating System Simulation), a novel framework that addresses these challenges by integrating code generation with a dynamic context management system. MOSS ensures consistency and adaptability by using a mechanism that maintains the Python context across interactions, including isolation of local variables and preservation of runtime integrity. At its core, the framework employs an Inversion of Control (IoC) container in conjunction with decorators to enforce the least knowledge principle, allowing agents to focus on abstract interfaces rather than concrete implementations. This facilitates seamless integration of new tools and libraries, enables runtime instance replacement, and reduces prompt complexity, providing a "what you see is what you get" environment for the agent.   

Through a series of case studies, we show how this framework can enhance the efficiency and capabilities of agent development and highlight its advantages in moving towards Turing-complete agents capable of evolving through code.
\end{abstract}

\begin{center}
    $\vcenter{\hbox{\includegraphics[height=1em]{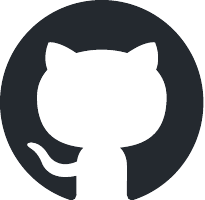}}}$
    \quad
    \url{https://github.com/ghost-in-moss/GhostOS}
\end{center}

\section{Introduction}

The development of AI agents\cite{dorri2018multi, wang2024survey} has seen rapid advancement in recent years, particularly with the integration of large language models (LLMs)\cite{brown2020language, touvron2023llama, achiam2023gpt, hui2024qwen2} into various applications and systems. In particular, AI agents, systems that can integrated tools\cite{qin2023toolllm, patil2023gorilla, tang2023toolalpaca} and interact with external environment, have recently received ever-increasing research focus. These agents are progressively tackling more complex tasks, ranging from software development\cite{yang2024swe, ma2024understand, zhang2024autocoderover} and web navigation\cite{zhou2023webarena}, even conducting scientific research\cite{lu2024ai, tyser2024ai}. As the field progresses, there is a growing interest in creating agents that can move beyond understanding and generating structured languages like JSON or domain-specific languages (DSLs)\cite{bassamzadeh2024plan}. The next frontier involves agents capable of using code as a medium for planning, reasoning, and dynamically creating tools, leading to the emergence of truly adaptive, code-driven AI agents.

AI agents driven by direct code generation\cite{ishibashi2024self} and execution mark the next evolution beyond JSON schema tools, offering Turing-complete capabilities and the potential for code-driven evolution. By using code to interact with tools and other agents, they enable a system where meta-agents—agents equipped with development tools—can expand their capabilities autonomously through code generation. This code-driven approach, while powerful, presents challenges in maintaining execution consistency, managing dynamic contexts, and isolating variables. In this paradigm, code-driven interaction is the method, meta-agents are the path, and adaptable, evolving behavior is the goal.   

Building AI agents capable of achieving Turing completeness and evolving through code is a major frontier in artificial intelligence research. Early efforts, such as Program-aided Language Models (PAL)\cite{gao2023pal} and CodeAct \cite{wang2024executable}, established the potential for large language models (LLMs) to automate code generation and execution, allowing agents to interact with environments by writing and running code. However, these methods primarily focus on single-turn interactions, limiting the agent’s ability to adapt to complex runtime contexts across multi-turn engagements.   

Subsequent approaches, like OpenDevin\cite{wang2024opendevin} and MindSearch\cite{chen2024mindsearch}, advanced multi-agent interactions and task planning, but still struggle with maintaining consistent execution contexts across multiple interactions—a key requirement for Turing completeness. For instance, MindSearch employs a multi-agent system for complex web retrieval but lacks dynamic adaptability in multi-turn interactions. Similarly, OpenDevin provides a sandboxed environment for agent interactions, yet relies on event streams, which lead to runtime inconsistencies and isolated execution contexts. Automated Design of Agentic Systems (ADAS)\cite{hu2024automated} introduced a meta-agent that iteratively programs new agents but also fails to maintain consistent code contexts across multi-turn tasks, limiting its autonomy and flexibility.   

Efforts like Diversity Empowered Intelligence (DEI)\cite{zhang2024diversity} form multi-agent ensembles to leverage diverse strengths, but do not achieve true Turing completeness as a single, evolving agent. Likewise, the "Literate Programming in the LLM Era" approach\cite{shi2024natural} synchronizes code with natural language for better understanding and maintenance, but focuses on aiding developers, not autonomous agent evolution.   

The central challenge in achieving Turing-complete AI agents is maintaining consistency between code and runtime context over multiple interactions. Current LLM-based systems often rely on single-shot code generation, depending on the LLM’s memory for subsequent turns. This leads to inefficiencies, as it fails to preserve the execution context or isolate variables between steps. Consequently, AI agents struggle with complex tasks requiring multi-step execution and dynamic adaptation. Moreover, many existing systems depend on manual protocol development in sandbox environments, limiting their potential for adaptive evolution.   

In this paper, we introduce MOSS (llM-oriented Operating System Simulation), a novel framework that addresses these challenges by integrating code generation with dynamic context management. Unlike previous methods reliant on event streams, sandboxed execution, or natural language summaries, MOSS ensures consistency by maintaining the Python runtime context across interactions. It isolates local variables and preserves runtime integrity, creating an environment where agents can expand their capabilities over time by generating code.   

MOSS aims to advance the field of AI agents in the following ways:   

1. Achieving True Turing Completeness: By ensuring consistency in code execution and context management across multi-turn interactions, MOSS enables AI agents to autonomously handle complex, multi-step tasks. This consistency allows agents to maintain execution context and isolate local variables, ensuring reliable task execution and dynamic adaptability in evolving environments.

2. Enabling Code-Driven Evolution and Extensibility: MOSS provides a framework that allows users or meta-agents to extend the functionality of AI agents through meta-coding techniques. By facilitating the integration of new tools and domain-specific agents at runtime, MOSS fosters continuous improvement within AI agents, enabling them to expand their capabilities through code.

\section{Method}

\subsection{Overview of the MOSS Framework}

MOSS (llM-oriented Operating System Simulation) is an essential component of the broader agent project GhostOS. GhostOS is designed to support advanced agent capabilities such as multi-task orchestration, environment interaction, and embodied control. MOSS specifically focuses on enabling AI agents to generate and execute Python code dynamically, while managing runtime contexts to ensure consistency across multi-turn interactions.   

MOSS operates on the following key principles:   

\paragraph{Python Context Reflection} It dynamically reflects the structure of Python modules, including variables, functions, and class definitions, into prompts. This allows the LLM to generate code with full awareness of the current execution context, ensuring the code is scoped accurately and appropriately.   

\paragraph{Dependency Injection via IoC} MOSS uses an Inversion of Control (IoC) container for injecting dependencies into the runtime. This enables the agent to integrate dynamically instantiated libraries, such as multi-task schedulers, chain-of-thought planners\cite{wei2022chain, li2024chain, zhang2024diagram}, environment sensors, and embodied APIs. By interacting through abstract interfaces, the agent remains adaptable to changing requirements without needing to modify its core logic.   

\paragraph{State Preservation Across Multi-Turn Interactions} The framework maintains a consistent execution context, preserving the state of variables and modules across multiple turns. Each interaction is isolated in its own frame, ensuring that local variables do not leak into other tasks, while global context is inherited as needed to support long-term task management.   

\paragraph{Execution of LLM-Generated Code} After generating Python code, MOSS executes it within the Python runtime. The results, along with runtime data, are integrated into the agent’s thought history, providing real-time feedback that informs subsequent decision-making and task execution.   

Beyond these core operational components, MOSS introduces a powerful mechanism for handling complex multi-task scenarios and long chains of thought. To manage this, the framework includes advanced scheduling and orchestration features within specialized libraries, which are accessible to the LLM. This allows agents to efficiently break down and execute intricate, multi-step tasks while maintaining overall coherence and adaptability.   

By integrating code generation, dependency injection, and dynamic context management, MOSS provides the foundation for developing Turing-complete AI agents. These agents can handle complex, multi-step tasks in an isolated and context-aware manner, with the potential for adapting and evolving their capabilities through code over time.

\subsection{Lifecycle of executing a task with MOSS}

The lifecycle of a MOSS agent involves managing complex, multi-turn tasks through a series of steps, organized into \textbf{Threads} and executed within \textbf{Frame} like mechanism. Each Thread represents a sequence of interactions, and each Frame manages the execution context for a specific step or set of steps. When the agent encounters an \textbf{AIFunc} or \textbf{Thought} in a higher-level frame (e.g., frame m), these intelligent units only have a definition but lack a body.

MOSS uses a recursive process to handle this. The \textbf{MOSS Compiler} steps in to compile the surrounding code and definitions, transforming them into a temporary Python module. This module provides the necessary execution context, including variables, dependencies, and AIFunc/Thought definitions, while keeping the environment isolated and consistent. The \textbf{MOSS Runtime} then processes this temporary module, converting it into a prompt for the large language model (LLM). The LLM generates the missing body, typically as a `main` function, to complete the task at hand.

After the code is generated, MOSS executes the new code within the current frame’s isolated context. If this execution involves further \textbf{AIFunc} or \textbf{Thought} calls, the agent drills down into a new frame (e.g., frame m+1), and the process repeats. This recursion allows MOSS to manage multi-step, complex tasks in a structured, isolated manner, ensuring each interaction's integrity through its \textbf{IoC container} and \textbf{PyContext}. Once the task in a frame is complete, the result is passed back to the parent frame, allowing the agent to continue processing seamlessly from where it left off.

This iterative cycle of code generation, execution, and feedback allows MOSS agents to manage dynamic, multi-turn interactions while preserving runtime context and achieving Turing-complete behavior.

The Figure \ref{fig:moss_lifecycle} illustrates the lifecycle of a MOSS agent as it manages multi-turn tasks, dynamically generating and executing code, maintaining consistency across steps, and ensuring isolated, context-aware interactions:

\begin{figure}
  \centering
  \includegraphics[width=1\textwidth]{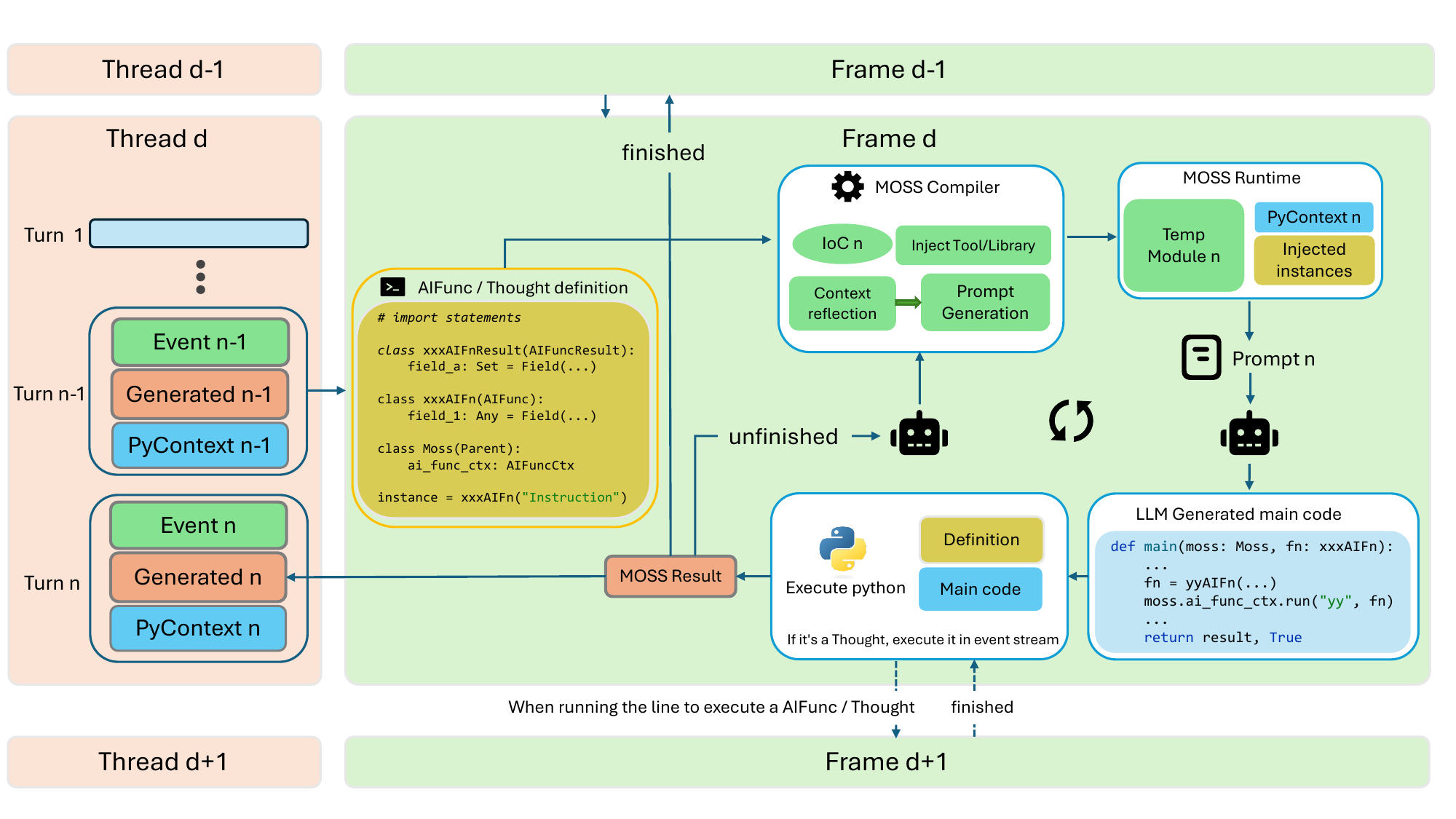}
  \caption{This diagram depicts the lifecycle of a MOSS agent as it handles complex, multi-turn tasks by dynamically generating and executing code. Each \textbf{Thread} represents a sequence of steps (\textbf{turns}) executed within \textbf{Frames}. Conceptually similar to function call stack frames in programming languages, each \textbf{Frame} represents an execution context that maintains its own dependencies and state. When the agent encounters an \textbf{AIFunc} or \textbf{Thought} in a higher-level frame (e.g., frame m), these components only have a definition but lack a body. MOSS then "drills down" into a new \textbf{Frame} (e.g., frame m+1), where the \textbf{MOSS Compiler} converts the surrounding code and definitions into a temporary Python module. The \textbf{MOSS Runtime} takes over, transforming this module into a prompt, which provides the LLM with the necessary context, dependencies, and AIFunc/Thought definitions. The LLM then writes the missing body (often a `main` function) to complete the function. After executing this new code, the result is passed back to the original frame, allowing the agent to continue processing from where it left off. Each frame maintains its own \textbf{IoC container} for dependency injection and manages its execution context independently through \textbf{PyContext}, ensuring consistency across multi-turn interactions. This iterative process of code generation, execution, and feedback enables MOSS to handle adaptive, multi-step tasks, achieving Turing-complete behavior.}
  \label{fig:moss_lifecycle}
\end{figure}

\subsection{Context Consistency}
Ensuring consistency across multi-turn interactions in MOSS starts with isolating each round of execution. Every interaction is compiled and executed in its own temporary ModuleType container, preventing any unintended side effects or variable leakage between rounds. This execution isolation ensures that each interaction operates in a clean, independent environment, free from interference by previous or subsequent steps.    

To maintain continuity, MOSS captures the side effects of each interaction through a serializable PyContext object. This PyContext holds local variable modifications and runtime dependencies, allowing relevant state information to persist across multiple turns. By doing this, MOSS ensures that while local variables remain isolated, essential context is preserved and carried forward, enabling the agent to dynamically adapt and evolve without losing track of its execution history.   

\subsection{Leveraging IoC for Runtime Flexibility and Abstraction}
In MOSS, the Inversion of Control (IoC) container plays a crucial role in decoupling the abstract interfaces that define agent capabilities from their runtime-specific implementations. This separation allows agents to operate within a dynamic environment where specific tools and functionalities are instantiated only when needed, without the LLM requiring knowledge of their internal workings. The IoC container enables MOSS to provide consistent access to these capabilities via interfaces, while abstracting away the complexity of how they are instantiated or modified at runtime.   
The primary advantage of IoC in MOSS is its ability to inject dependencies dynamically at runtime, following the least knowledge principle. This ensures that the LLM can focus on leveraging abstract interfaces—such as task schedulers, environment sensors, or APIs—without needing to understand or instantiate their underlying implementations. For example, an LLM may generate code to interact with an interface for file management or debugging, while MOSS handles the actual instantiation of these tools based on the runtime context, allowing for seamless integration of diverse libraries and tools as needed.   

This design pattern also aligns with the Liskov Substitution Principle\cite{martin2003agile}, as it allows different implementations of the same interface to be substituted without affecting the LLM's understanding of the task. This makes it possible to mock these interfaces during testing, allowing developers to refine prompts and verify the agent’s use of the interface before integrating actual implementations. By doing so, developers can optimize LLM behavior under controlled conditions and later replace mocks with the real dependencies without changing the LLM’s core logic or prompts.    
In essence, IoC in MOSS allows agents to remain adaptable and modular, facilitating the smooth integration of new tools and capabilities at runtime. This also improves testing and prompt refinement, as developers can easily experiment with abstracted interfaces before committing to full implementations. By decoupling runtime-specific details from high-level task execution, IoC ensures that MOSS agents are more flexible and maintainable.

\subsection{Intelligent Units: AIFunc and Thought}
MOSS operates as a lightweight framework that facilitates the use of external intelligent units, such as AIFunc and Thought, although they do not originate from MOSS itself. MOSS provides the necessary infrastructure for running these intelligent units in a structured and context-consistent manner, serving as an llM-oriented OS simulation that can be integrated into other agent frameworks. These intelligent units leverage MOSS for their execution but are designed to extend agent capabilities by allowing multi-turn code generation and interaction with complex runtime environments.

\paragraph{AIFunc}  AIFunc represents synchronous intelligent functions. An AIFunc appears as a typical Python function that can be invoked with predefined parameters. However, in practice, its execution involves single-turn interactions where it generates and executes code across multiple steps, often driven by real-time feedback. Each AIFunc operates within its own execution frame, ensuring the isolation of local variables while maintaining access to a shared global context. By nesting AIFuncs or calling them from within Thoughts, agents can handle complex, multi-layered tasks through code that adapts dynamically to the current execution context.  

\paragraph{Thought}  Thought represents asynchronous, multi-turn intelligent processes. Unlike AIFunc, which focuses on single interactions, Thought engages in ongoing reasoning, adapting strategies dynamically based on feedback from the environment. Thoughts run asynchronously and can manage multiple complex tasks in parallel, using tools and code execution provided by MOSS. Thought functions as a domain-specific intelligent unit, solving specialized problems while sharing the same meta-prompt as other agents. This enables more complex, long-term task orchestration and problem-solving.   

\subsection{Code-Driven Evolving}

MOSS establishes a code-driven framework as the foundation for AI systems capable of evolving through code. This paradigm shifts the focus from traditional development protocols to a system where code itself drives the creation, interaction, and evolution of agents. By treating code as the central mechanism, MOSS maximizes flexibility and adaptability during agent development, allowing for continuous improvements. This code-centric approach is a significant step toward self-evolving systems, where agents can dynamically adjust and enhance their capabilities based on new code inputs and interactions.

\begin{figure}
  \centering
  \includegraphics[width=1\textwidth]{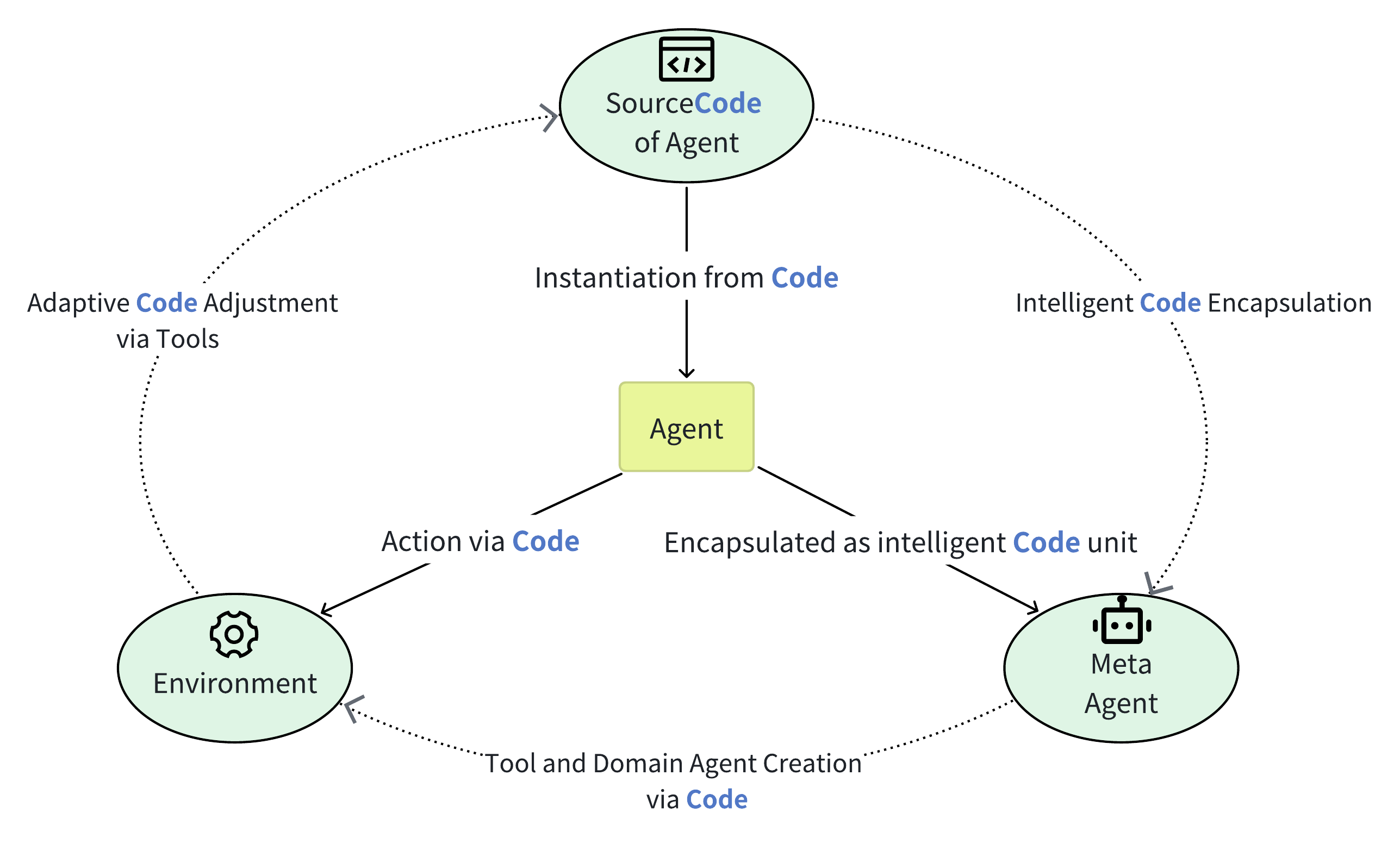}
  \caption{This diagram illustrates the code-driven evolving mechanism within MOSS. The system enables dynamic interactions between meta-agents, the source code, and the environment. Tools within the environment continuously modify agent source code, which is then used by the Meta-Agent to create encapsulated intelligent code units (e.g., AIFunc or Thought). The Meta-Agent uses these units to develop new tools and agents, thereby perpetuating a cycle of code-driven evolution.}
  \label{fig:evolve}
\end{figure}

\paragraph{Code-Driven Tool/Library Integration} 
MOSS allows agents to interact with tools and libraries through dynamically generated Python code, reducing the need for manual protocol development. While tools still require defined interfaces, the goal is for agents to directly read and understand code repositories, integrating them as tools. Future libraries may evolve into interface-based code repositories, where agents can autonomously bind implementations using dependency injection. This minimizes integration overhead and allows embodied agents to dynamically develop drivers and tool controls through code.

\paragraph{Code-Defined Agents}
MOSS enables agents to be fully defined and modified by code, removing the constraints of domain-specific languages (DSLs). This approach gives agents maximum flexibility and allows for dynamic adjustments. By using popular programming language like Python, developers or LLMs can easily define, reuse, and extend agents in familiar environments. Developers or LLMs can modify agent's behavior through code, leading to a more scalable and adaptive development process compared to rigid DSL-based systems.

\paragraph{Code-Orchestrated Agent Interaction}
Agents in MOSS interact with each other through code, treating one another as callable components. This facilitates dynamic orchestration, where agents can retrieve, invoke, or coordinate tasks through code. Instead of relying on fixed interaction protocols, MOSS allows agents to collaborate and adapt in real-time, forming a more flexible and interconnected agent ecosystem.

\paragraph{Code-Driven Meta-Agent Definition}
In MOSS, meta-agents dynamically create new agents and tools through code, not only managing but also creating more complex workflows. By generating code to build or extend agents, meta-agents continuously expand the system's capabilities. This recursive process allows for the development of higher-order agents capable of addressing increasingly intricate tasks. Crucially, meta-agents are not limited to their current generation—they can iteratively develop more advanced agents, continually pushing the boundaries of complexity and functionality in the system.

\paragraph{Closing the Code-Driven Evolutionary Loop}
MOSS creates a closed evolutionary loop by integrating code-based tool/library integration, agent definition, agent interaction, and meta-agent-driven orchestration. As seen in Figure \ref{fig:evolve}, the system allows tools in the environment to modify agents’ source code dynamically, which the Meta-Agent encapsulates and uses to create new intelligent units. These intelligent units are then employed by the Meta-Agent to develop new tools and domain-specific agents, further enhancing the system’s capabilities. While not yet fully autonomous, MOSS establishes the foundation for future systems capable of evolving on their own, driven entirely by code.

\section{Case Studies}

\subsection{Create a Tool through Code Manipulation}
In this case, we explore how MOSS enables LLMs to edit existing code dynamically and generate a new tool by interacting with the code via a custom tool called ModuleEditor. ModuleEditor provides an interface that allows agents to read, modify, and append Python code in a target module, and it integrates into the MOSS framework through the IoC container. In this example, the agent is tasked with generating the implementation of a caching tool (MockCache), dynamically editing the module by adding the necessary class definitions.

This showcases the ability of MOSS to allow agents to utilize dynamically injected tools to edit and expand code autonomously.

\paragraph{Setup}
The MOSS framework dynamically reflects the ModuleEditor interface into the LLM’s runtime context via dependency injection using the IoC container. The MOSS file defines a Moss class, which integrates ModuleEditor as an attribute. This attribute is exposed to the LLM as part of the prompt, allowing the agent to directly invoke ModuleEditor's methods during task execution.    

In the MOSS file, ModuleEditor is injected as a tool, and the agent is tasked with replacing the MockCache class definition within a Python module. The agent achieves this by calling the replace\_attr method of ModuleEditor to modify the code in place. The task prompt is augmented by MOSS to ensure the LLM has full knowledge of the ModuleEditor methods, ensuring smooth interaction with the Python module.   

\paragraph{Thought Definition: PyModuleEditorThought}
To guide the agent in its task, developer or a meta-agent can define a Thought that provides the agent with instructions on how to modify the module. The agent is tasked with implementing the MockCache class in the target module by leveraging the ModuleEditor.   

\begin{lstlisting}
DEFAULT_PY_MODULE_EDITOR_INSTRUCTION = """
# Instruction

Your task is helping the user to read/understand/edit a python module file. 
The target module you are handling is `{modulename}`, the source code with line num are: 

```python
{target_source}
```
"""
\end{lstlisting}

\paragraph{Results and Analysis} 

The agent successfully implements the MockCache class by generating the appropriate Python code and using the replace\_attr method of ModuleEditor to replace the placeholder class definition with the new implementation. The result demonstrates the agent’s ability to autonomously modify code through tool integration, allowing for dynamic code evolution. By employing the MOSS framework, the agent ensures that the newly generated code is correctly inserted into the target module, preserving the execution context and maintaining isolation between tasks.   

The successful implementation of MockCache highlights the importance of MOSS files in defining the agent’s runtime environment. The MOSS file plays a crucial role in shaping the prompt seen by the LLM, which includes the necessary tool interfaces and system-wide configuration. The tool-driven approach is facilitated by MOSS's IoC container, allowing for seamless interaction between the agent and external libraries.     

More details of pipeline for this case study can be found in \hyperref[app:tool_integration]{Appendix A}.

\subsection{Asynchronous Multi-Task Management}

In this case, we show how MOSS enables asynchronous multi-task scheduling through the integration of the MultiTask library, efficiently managing and executing multiple file-editing tasks concurrently. The scenario involves editing a directory of .py files to translate Chinese comments into English. MOSS's capability to orchestrate multiple \textbf{FileEditorThought} instances allows it to dispatch and manage these tasks asynchronously, showcasing its ability to handle complex, multi-step operations in parallel.   

\paragraph{Setup}  
The DirectoryEditorThought is defined to manage a directory of files, while the FileEditorThought focuses on editing individual files. Through the MOSS file, three key attributes are injected into the MOSS class: Replier, MultiTask, and DirectoryEditor. These provide the core functionality to handle directory editing and multi-tasking. The user request, which could be initiated by a meta-agent, involves translating any Chinese comments found in the .py files within the code\_edits directory into English.   

The MOSS file plays a crucial role here by ensuring that all necessary tools (such as MultiTask and DirectoryEditor) are integrated through dependency injection in the IoC container. It also ensures that the prompt seen by the LLM provides a WYSIWYG (What You See Is What You Get) environment, where the LLM sees a global system prompt to generate a main(moss: MOSS) method with the injected tools made available in the MOSS class.   

During the task execution, the DirectoryEditorThought will list all .py files in the target directory, then dispatch tasks to individual FileEditorThought instances via MultiTask. Each FileEditorThought receives a task to open a file, read the contents, and translate the comments from Chinese to English. These tasks are processed asynchronously.   

\paragraph{Results and Analysis}
Once the tasks are dispatched, MultiTask asynchronously schedules and manages the execution of each FileEditorThought. For each file, the FileEditorThought reads the file’s contents, identifies any Chinese comments, replaces them with English translations, and then completes the task.   

After all tasks are completed, DirectoryEditorThought receives the task status messages, consolidating the results and notifying the user of task completion. The final messages show that all tasks, including translating the comments in multiple Python files, were executed successfully in parallel.   

This case highlights MOSS's ability to manage multiple tasks concurrently through code-driven interaction with tools, emphasizing the role of the MOSS file in ensuring that the LLM operates with a clear, WYSIWYG interface. It also showcases how MOSS handles dynamic task management, context consistency, and dependency injection to streamline the process of handling complex, multi-step tasks.   

Detailed internal process and results of this case study can be found in \hyperref[app:async_multi_task]{Appendix B}.

\subsection{Debugging an Issue in Code Repository}
In this case study, we demonstrate the capabilities of the MOSS framework by debugging a specific issue in the Django framework's messages module. 
This case is sampled from SWEBench-lite\cite{jimenez2023swe}. The issue involved incorrect serialization of 'extra\_tags' when it was an empty string, leading to unexpected behavior in the application. Using MOSS, we implemented a systematic approach to localize the bug efficiently.

\paragraph{Setup}
The debugging process began by setting up the environment for the Django repository, specifically focusing on the commit 7c4f3965098baad2396e24501e09237425a7bd6f. We defined a series of AI functions (AIFuncs) to manage the workflow. The initial step involved gathering task metadata and preparing the repository for debugging. Subsequently, the framework utilized a search strategy to locate files related to the 'extra\_tags' issue and explored their contents. This involved multiple iterations of exploration and exploitation to pinpoint the culprit file effectively.

\paragraph{Results and Analysis}
Through the execution of the MOSS framework, we successfully identified the root cause of the issue within the django/contrib/messages/storage/cookie.py file. The exploration strategy yielded a confidence percentage exceeding our defined threshold, validating the effectiveness of our approach. The overall process demonstrated the adaptability and efficiency of MOSS, showcasing its ability to manage complex debugging tasks while maintaining context and facilitating code-driven evolution. By defining several AIFuncs without function bodies and importing necessary tools, the agent dynamically generated and executed workflows, proving our achievement of Python context consistency across multiple interactions. This highlights an important cognitive aspect: the primary driver between multiple turns is pure code. Even without explicitly writing a workflow, the defined AIFuncs executed as if they were part of a complete workflow, with the connections between workflow nodes being Turing complete. The result illustrates the potential of MOSS in orchestrating complex multi-step tasks and managing Python context across multiple turns. 

Detailed information of this case study can be found in \hyperref[app:swe_bench_lite]{Appendix C}.

\section{Discussion}

\subsection{Security Considerations}
One of the key design choices in MOSS is its operation within the local Python process rather than a sandboxed environment. While this decision facilitates direct integration and interaction with the IoC container, enhancing execution efficiency and context management, it raises potential security concerns. Unlike sandboxed environments, where code execution is isolated and risks are mitigated, the local process approach requires careful management to prevent unintended side effects or security vulnerabilities. Future work will focus on improving the safety of the local execution environment, implementing more robust safeguards to prevent harmful code execution, and ensuring the integrity of the system. We invite the community to collaborate on developing best practices and contributing to the security framework of MOSS.

\subsection{Enhancing Thought and AIFunc with Advanced Models}
MOSS introduces Thought and AIFunc as intelligent units for executing complex, multi-turn tasks. While these units already offer significant capabilities, their potential can be further unlocked through advanced language models like the GPT-4 series, known for their strong chain-of-thought (CoT) reasoning abilities. Such models can improve the coherence and effectiveness of Thought processes and enhance the accuracy of AIFunc operations by generating more context-aware and purpose-driven code. By leveraging the inherent reasoning strengths of these models, MOSS can execute complex sequences of actions more effectively, making the framework even more suitable for tasks that require adaptive planning, exploration, and execution.

\subsection{The Need for Amplifiers and Integration Frameworks}

Even the most advanced AI models, like GPT-4\cite{achiam2023gpt}, benefit significantly from amplifiers—tools and frameworks that augment their problem-solving capabilities\cite{chen2023agentverse, wang2023mint}. Just as humans rely on IDEs and methodologies like divide-and-conquer to tackle complex software development tasks, AI agents require a well-integrated framework to interact with their environment and manage complex workflows. MOSS serves as this amplifier, providing an integrated code-generation and context-management system that allows AI agents to decompose intricate tasks into manageable units. By systematically interacting with their environment through code, agents can achieve higher-order functionalities that transcend the limitations of any single model. This framework enables the development and deployment of sophisticated AI systems that can be debugged, monitored, and guided more effectively by humans, ensuring reliable and controllable AI behavior.

\subsection{Enhancing Debuggability and Control in Complex AI Systems}
A core advantage of MOSS's design philosophy is the emphasis on code-driven interaction with the environment, which offers a higher degree of transparency and control compared to black-box AI systems. By adopting a divide-and-conquer approach and focusing on generating and executing code, MOSS allows developers to debug complex AI systems in a manner similar to traditional software debugging. This structured approach provides a clear trace of the agent's decision-making process, enabling developers to intervene, modify, and guide the AI system at critical junctures. By integrating human expertise into the evolution of AI agents through direct code modification, MOSS offers a path toward creating more controllable and interpretable AI systems that can adapt and evolve incrementally while remaining aligned with human objectives.

\section{Conclusion}
MOSS represents a significant advancement in the development of Turing-complete AI agents. By integrating code generation with dynamic context management, it addresses the challenges of maintaining consistency between code and runtime context across multi-turn interactions. The framework's use of an IoC container and runtime instance replacement facilitates the seamless integration of new tools and libraries, enabling agents to expand their capabilities over time. Furthermore, MOSS's structured approach to executing complex, multi-step tasks through isolated execution frames and intelligent units like AIFunc and Thought empowers agents to adapt and tackle increasingly intricate challenges.

The framework's design also emphasizes the importance of transparency and control in AI systems, offering developers a robust platform to debug and guide agents using familiar programming paradigms. While challenges such as execution security remain, MOSS lays the groundwork for future research and development in creating more adaptable, interpretable, and secure AI agents. The integration of advanced language models and the continued refinement of the framework promise to further enhance the capabilities of AI agents, moving towards the vision of more autonomous and adaptive intelligent systems.

\bibliographystyle{unsrt}  
\bibliography{references}  
\nocite{*}  

\appendix
\newpage  
\noindent\textbf{Appendix Contents} 
\section{Create a Tool through Code Manipulation} \label{app:tool_integration}
\subsection{Tool: ModuleEditor}
We implemented a Python class library, ModuleEditor(you can find it in \url{https://github.com/ghost-in-moss/GhostOS/tree/main/ghostos/demo/src/examples/code_edits/py_editor_thought_test.py}), designed to modify Python module files. This class offers various methods to interact with and edit the source code of a target module. The abstract class ModuleEditor is defined as follows:

\begin{lstlisting}
class ModuleEditor(ABC):
    """
    module editor that instance from a target module.
    can edit the target module's contents.
    """

    @abstractmethod
    def filepath(self) -> str:
        """
        :return: filepath of the target module
        """
        pass

    @abstractmethod
    def modulename(self) -> str:
        """
        :return: module name of the module that this editor is editing
        """
        pass

    @abstractmethod
    def read_source(
            self,
            show_line_num: bool = True,
            start_line: int = 0,
            end_line: int = -1,
    ) -> str:
        """
        read source code from this module
        :param show_line_num: show line number at the end of each line such as # 44
        :param start_line: start line number
        :param end_line: end line number, if < 0, means end line number
        :return: source code
        """
        pass

    @abstractmethod
    def read_source_of_imported(
            self,
            attr_name: str,
            detail: bool = False,
    ) -> str:
        """
        read a imported attribute's source code.
        :param attr_name: the attribute's name of the target in this module
        :param detail: if True, show source code; if False, show signature only
        :return: full source code, or a simple string describe it.
        """
        pass

    @abstractmethod
    def replace(
            self,
            target_str: str,
            replace_str: str,
            count: int = -1
    ) -> bool:
        """
        replace the source code of this module by replace a specific string
        :param target_str: target string in the source code
        :param replace_str: replacement
        :param count: if -1, replace all occurrences of replace_str, else only replace occurrences count times.
        :return: if not ok, means target string is missing
        """
        pass

    @abstractmethod
    def replace_block(
            self,
            start_line: int,
            end_line: int,
            replace_str: str,
    ) -> str:
        """
        replace a block of source code
        :param start_line: the start line number of the block.
        :param end_line: the end line number of the block, included.
        :param replace_str: replacement
        :return: the replaced source code, if empty, means target block is missing
        """
        pass

    @abstractmethod
    def replace_attr(
            self,
            attr_name: str,
            replace_str: str,
    ) -> str:
        """
        replace a module attribute's source code.
        the target attribute shall be a class or a function.
        :param attr_name: name of the target attribute of this module. It MUST be defined in this module, not imported.
        :param replace_str: new source code
        :return: the replaced source code. if empty, means target attribute is missing
        """
        pass

    @abstractmethod
    def append(self, source: str) -> None:
        """
        append source code to this module.
        :param source: the source code of class / function / assignment
        """
        pass

    @abstractmethod
    def insert(self, source: str, line_num: int) -> None:
        """
        insert source code to this module at line number.
        remember following the python code format pattern.
        :param source: the inserting code, such like from ... import ... or others.
        :param line_num: the start line of the insertion
        """
        pass
\end{lstlisting}

\subsection{MOSS File for ModuleEditor Integration}
The MOSS file defines the Moss class, which includes an injected ModuleEditor tool that the agent can use to interact with the module's source code. The prompt visible to the LLM is nearly identical to the actual source code, ensuring a "what you see is what you get" (WYSIWYG) experience. The LLM uses this interface to generate Python code that manipulates the target module via the ModuleEditor methods.
\begin{lstlisting}
from typing import Optional
from ghostos.core.ghosts import Operator, Replier
from ghostos.core.moss import Moss as Parent
from ghostos.libraries.py_editor import ModuleEditor

class Moss(Parent):
    """
    Moss that equipped with ModuleEditor
    """
    editor: ModuleEditor
    """ the editor about target module """

    replier: Replier
    """ with replier you can send reply in main function"""

if __name__ == "__example__":
    def example_append_code_at_main(moss: Moss) -> Optional[Operator]:
        """
        this example is about you need to add codes to the target module.
        """
        # write the target code as string variable
        code = """
def plus(a, b):
    return a + b
"""
        # using editor api to add code
        moss.editor.append(code)
        # return none means if print anything, observe them and think again. otherwise do default action awaits.
        # NEVER CONFUSE the MOSS interface and the target module.
        # 1. MOSS interface providing you with a python interface to using many libraries, instead of JSON Schema tools.
        # 2. Target module is what you want to handle.
        return None
\end{lstlisting}

\subsection{Thought for Code Modification}
The PyModuleEditorThought class defines the context for the agent to use the ModuleEditor to manipulate Python modules. This thought process drives the agent's actions when interacting with the module's source code.

\begin{lstlisting}
DEFAULT_PY_MODULE_EDITOR_INSTRUCTION = """
# Instruction

Your task is helping user to read / understand / edit a python module file. 
The target module you are handling is `{modulename}`, the source code with line num are: 

```python
{target_source}
```
With ModuleEditor that MOSS provided you can read / change it. 
Remember to print the result when you call some method of ModuleEditor, check the printed result if anything goes wrong.

You can update `{modulename}` 's code, use the ModuleEditor that MOSS provided to you. 
ModuleEditor provides multiple methods to update the source code, you need to write your code as a string, and use the methods.
"""

class PyModuleEditorThought(ModelThought):
    target_module: str = Field(description="target modulename")
    referencing: Dict = Field(
        default_factory=dict,
        description="references for python editor, key is import path, value is the prompt about it"
    )
    llm_api_name: str = Field(default="", description="specific llm api name")
    debug: bool = Field(default=False, description="debug mode, if true, moss code will send to user")
\end{lstlisting}

\subsection{New tool implementation via ModuleEditor}
In this example, we aim to implement a caching tool, `MockCache`, using the `ModuleEditor`. The task involves creating the class `MockCache` with various caching methods and replacing its placeholder definition in the target module using `replace\_attr`.
\begin{lstlisting}
class MockCache(Cache):
    """
    Mock for cache, expected to be implemented using dict.
    """

    def lock(self, key: str, overdue: int = 0) -> bool:
        pass

    def unlock(self, key: str) -> bool:
        pass

    def set(self, key: str, val: str, exp: int = 0) -> bool:
        pass

    def get(self, key: str) -> Optional[str]:
        pass

    def expire(self, key: str, exp: int) -> bool:
        pass

    def set_member(self, key: str, member: str, value: str) -> bool:
        pass

    def get_member(self, key: str, member: str) -> Optional[str]:
        pass

    def remove(self, *keys: str) -> int:
        pass


if __name__ == '__main__':
    from ghostos.prototypes.console import new_console_app
    from ghostos.thoughts import new_pymodule_editor_thought
    app = new_console_app(__file__, 4)
    app.run_thought(
        new_pymodule_editor_thought(__name__),
        instruction="implement the MockCache for me.",
        debug=True,
    )
\end{lstlisting}

\subsection{Generated Code for MockCache Replacement}
Here, the LLM-generated code implements the MockCache class, and the replace\_attr method is used to replace the placeholder MockCache class in the target module with this newly defined implementation.   

\begin{lstlisting}
def main(moss: Moss) -> Optional[Operator]:
    # The implementation for MockCache class
    mock_cache_code = """
class MockCache(Cache):
    def __init__(self):
        self.store = {}
        self.locks = {}

    def lock(self, key: str, overdue: int = 0) -> bool:
        if key in self.locks:
            return False
        self.locks[key] = True
        return True

    def unlock(self, key: str) -> bool:
        if key in self.locks:
            del self.locks[key]
            return True
        return False

    def set(self, key: str, val: str, exp: int = 0) -> bool:
        self.store[key] = val
        return True

    def get(self, key: str) -> Optional[str]:
        return self.store.get(key)

    def expire(self, key: str, exp: int) -> bool:
        # For simplicity, this mock implementation doesn't handle expiration.
        return True

    def set_member(self, key: str, member: str, value: str) -> bool:
        if key not in self.store:
            self.store[key] = {}
        if isinstance(self.store[key], dict):
            self.store[key][member] = value
            return True
        return False

    def get_member(self, key: str, member: str) -> Optional[str]:
        if key in self.store and isinstance(self.store[key], dict):
            return self.store[key].get(member)
        return None

    def remove(self, *keys: str) -> int:
        removed_count = 0
        for key in keys:
            if key in self.store:
                del self.store[key]
                removed_count += 1
        return removed_count
    """

    # Replace the MockCache class in the module
    moss.editor.replace_attr("MockCache", mock_cache_code)
    return moss.replier.finish("MockCache class has been implemented successfully.")
\end{lstlisting}

Through this process, we successfully implemented the MockCache class using the ModuleEditor tool, demonstrating how to leverage MOSS to integrate existing tools and generate new tools into the system.

\section{Asynchronous Multi-Task Management} \label{app:async_multi_task}
This appendix showcases how MOSS leverages the MultiTask library to manage multiple FileEditorThought instances concurrently, performing asynchronous task execution. In this example, MOSS is used to read a directory of .py files, and each file is edited by translating Chinese comments into English. This demonstrates the asynchronous task scheduling capabilities of MOSS. All codes can be found in \url{https://github.com/ghost-in-moss/GhostOS/blob/main/ghostos/demo/src/examples/code_edits/modify_directory_test.py}

\subsection{DirectoryEditorThought Definition}
The DirectoryEditorThought class is responsible for managing a directory of files. The thought can list files and invoke tasks on them. It integrates with MultiTask to dispatch and manage multiple tasks concurrently.

\begin{lstlisting}
# import statements

class DirectoryEditorThought(MagicMossThought):
    """
    Useful to manage all files in a directory
    """
    directory: str = Field(description="absolute path to the directory to be edited")
    debug: bool = Field(default=False, description="turn debugging on")
    llm_api_name: str = Field(default="", description="name of the llm api")


class Moss(Parent):
    """
    you are equipped with some tools helping you to manage the current directory.
    and the FileEditorThought are helpful to manage a single file.
    """

    replier: Replier

    multitask: MultiTask
    """useful to handle multiple tasks, such as manage several files by FileEditorThought."""

    dir_editor: DirectoryEditor
    """ 
    the editor of the current directory.
    you can read or edit the files by FileEditorThought.
    so don't make up anything, based on what you informed.
    """


# <moss>
# the codes between the moss xml marks are not visible to LLM

# using TYPE_CHECKING to avoid reflect invalid importing to prompt.
if TYPE_CHECKING:
    from ghostos.core.ghosts import Ghost
    from ghostos.core.session import Event, Session, MsgThread
    from ghostos.core.llms import LLMApi
    from ghostos.core.moss import MossCompiler

# Instruction 
# Your task are handling files in the current directory.
# Current directory is `{current_dir}`, recursive depth is `{depth}`.
# The files and subdirectories are listed below:

```
{list_info}
```

# You shall use dir_editor and FileEditorThought to fulfill the user's request. 

**Notices**
* the best way to handle single file is to use FileEditThought which will see the detail of the file.
* once you list absolute filepaths, do not join it with some directory prefix.
* do not imagine the content of the files.
"""
    instruction = temp.format(
        current_dir=thought.directory,
        depth=depth,
        list_info="\n".join(directory_info),
    )
    return instruction

# </moss>
\end{lstlisting}

\subsection{FileEditorThought Definition}
The FileEditorThought class is used to manage individual file edits. Each file is read, comments in Chinese are identified, and then replaced with their English translations.

\begin{lstlisting}
class FileEditorThought(ModelThought):
    """
    Useful to read, understand and modify a file with any request.
    """
    filepath: str = Field(description="absolute filepath the thought shall edit on")
    debug: bool = Field(default=False, description="if debug mode, generate relative thread file for debugging")
    llm_api_name: str = Field(default="", description="llm model name, if you don't know what you want, keep empty")
\end{lstlisting}

\subsection{MOSS File}
The MOSS file plays a critical role in injecting the necessary tools—Replier, MultiTask, and DirectoryEditor—into the MOSS class. These tools are available for task execution, ensuring that the LLM has access to the required methods and can generate task-oriented code.

\subsection{Task Execution}

The user gives the following instruction, which is processed by DirectoryEditorThought:   
"please checkout content of the `.py` files in code\_edits directory, 
and translate the comments in  chinese into english if you found them in the code."

The DirectoryEditorThought generates a plan to iterate through the .py files in the code\_edits directory. For each file, a FileEditorThought is created to handle the individual file-editing task. These tasks are dispatched to MultiTask, which runs them asynchronously.   
\begin{lstlisting}
    def main(moss: Moss) -> Operator:
    # List all .py files in the code_edits directory
    code_edits_files = moss.dir_editor.edit_dir("code_edits").list(pattern="*.py", absolute=True, formated=False, summary=False)

    # Create tasks for checking out the content of each file
    tasks = []
    for file in code_edits_files:
        task = (
            f"translate_comments_{file.split('/')[-1]}",
            f"Translate comments in {file} from Chinese to English",
            FileEditorThought(filepath=file),
            "Please read the file content and translate any comments in Chinese to English."
        )
        tasks.append(task)

    # Run the tasks
    return moss.multitask.wait_on_tasks(*tasks)
\end{lstlisting}

\subsection{Sub Task Execution}
When a FileEditorThought receives a task, it reads the file's content and translates the Chinese comments to English. Below is an example of the code executed to perform the task.

\begin{lstlisting}
def main(moss: Moss) -> Optional[Operator]:
    # Read the content of the file
    file_content = moss.editor.read(show_line_num=True)

    # Replace Chinese comments with English, some specific replace statements
    ...

    # Print the modified content
    print(moss.editor.read(show_line_num=True))

    # Finish the task and reply with the updated content
    return moss.replier.finish("Translated all Chinese comments to English.")
\end{lstlisting}

\subsection{Task Result}

Once all tasks have been executed asynchronously by MultiTask, DirectoryEditorThought receives feedback that the tasks are completed and notifies the user. Example task completion messages are as follows:

\begin{lstlisting}
- name: __system__
  role: assistant
  content: receive event wait_callback from task `translate_comments___init__.py`.
- name: GhostOSDemo
  role: assistant
  content: receive event finish_callback from task `translate_comments_modify_directory_test.py`.
- name: __system__
  role: assistant
  content: receive event finish_callback from task `translate_comments_file_editor_test.py`.
- name: GhostOSDemo
  role: assistant
  content: Translated all Chinese comments to English.
\end{lstlisting}

This simple demo illustrates MOSS's ability to manage and execute asynchronous multi-tasking using MultiTask while demonstrating how MOSS integrates tools and libraries through dependency injection. The system can handle much more complex tasks by adding external tools and Thought definitions, making it highly adaptable for more powerful applications.

\section{Debugging an Issue in Code Repository} \label{app:swe_bench_lite}

This is an example of MOSS performing root cause localization for an issue in a code repository using AIFunc. The target repository is \url{https://github.com/django/django}, with the base commit being 7c4f3965098baad2396e24501e09237425a7bd6f. The issue is described as follows:

\begin{lstlisting}[style=plaintextstyle]
Messages framework incorrectly serializes/deserializes extra_tags when it's an empty string
Description
	
When a message is serialised and then deserialised with any of the built in storage backends, then extra_tags=="" is converted to extra_tags==None. This is because MessageEncoder checks for the truthyness of extra_tags rather than checking it is not None.
To replicate this bug
>>> from django.conf import settings
>>> settings.configure() # Just to allow the following import
>>> from django.contrib.messages.storage.base import Message
>>> from django.contrib.messages.storage.cookie import MessageEncoder, MessageDecoder
>>> original_message = Message(10, "Here is a message", extra_tags="")
>>> encoded_message = MessageEncoder().encode(original_message)
>>> decoded_message = MessageDecoder().decode(encoded_message)
>>> original_message.extra_tags == ""
True
>>> decoded_message.extra_tags is None
True
Effect of the bug in application behaviour
This error occurred in the wild with a template tag similar to the following:
{% if x not in message.extra_tags %}
When the message was displayed as part of a redirect, it had been serialised and deserialized which meant that extra_tags was None instead of the empty string. This caused an error.
It's important to note that this bug affects all of the standard API (messages.debug, messages.info etc. all have a default value of extra_tags equal to "").
\end{lstlisting}

First, we define the following AIFunc, which will serve as the primary AIFunc, outlining the general execution flow of the tasks from a high-level perspective:

\subsection{Debug Localization AIFunc}

\begin{lstlisting}
# other import statements
from ghostos.core.moss import Moss as Parent
from evaluation.swe_bench_lite.ai_funcs.swe_task_manager import SWETaskAIFunc, SWEDebugTaskCtx
from evaluation.swe_bench_lite.ai_funcs.project_explorer import ExplorationProjectAIFunc

class DebugAgentFn(AIFunc):
    """
    AIFunc that act like an agent
    """
    request: str = Field(default="", description="raw request for the agent")


class DebugAgentFnResult(AIFuncResult):
    """
    the result that follow the agent request
    """
    issue_culprit_file_parts: Set[CulpritFilePart] = Field(default=[], description="the file parts that caused the issue")
    err: Optional[str] = Field(default=None, description="error message")


class Moss(Parent): 

    ai_func_ctx: AIFuncCtx
    """useful to run AIFunc"""

# <moss>
def __aifunc_instruction__(fn: DebugAgentFn) -> str:
    return fn.request

example = DebugAgentFn(
    request="Your task is localization issue files in a repository. "
            "First get the information of the swe bench task"
            "Then using prepare the environment to debug the repository. "
            "Then localize the file caused the issue (not mock, it might be a Localization(exploration and exploitation) AIFunc). "
            "If you realize some steps needs to utilizing AI to plan or implementation, utilize the AIFunc. "
            "Task json file path: /home/llm/swe_bench_lite/django_15347.json  "
            "workspace path: /home/llm/workspace/django"
            # "You can create AIFunc by definition class outside of the `def main(moss)`"
)

# </moss>
\end{lstlisting}

Similarly, we will create AIFuncs for task retrieval and repository exploration, which will be used to obtain task information and prepare the repository environment. Each AIFunc will only have method signatures without implementations, and they will be imported into the DebugLocalizationAIFunc.

\subsection{SWETask AIFunc}

\begin{lstlisting}
# other imports
from ghostos.core.aifunc import AIFunc, AIFuncResult
from ghostos.core.moss.decorators import cls_source_code

pattern = r'(\w+)\s\(([\w\.]+)\)'

compiled_pattern = re.compile(pattern)

def extract_info(text):
    match = compiled_pattern.search(text)
    if match:
        method_name, class_path = match.groups()
        return method_name, class_path
    return None

class UnitTestInfo(BaseModel):
    test_method_name: str = Field(..., description="The name of the test method")
    test_class_name: str = Field(..., description="The name of the test class from repository content root")

SEP = "\n====================================\n"

@cls_source_code()
class SWEDebugTaskCtx(BaseModel):
    workspace_path: str = Field(..., description="The path of the root workspace")
    repo: str = Field(..., description="The target repository to debug")
    instance_id: str = Field(..., description="The id of the debug task")
    base_commit: str = Field(..., description="The base commit of the repository to debug")
    issue_info: str = Field(..., description="The issue statement of the bug")
    supplementary_issue_info: str = Field(..., description="The discussion and supplementary text of the bug")
    passed_tests: List[UnitTestInfo] = Field(..., description="The list of passed unit tests before the fix")

def _get_method_name_and_class_from_str(s: str) -> (str, str):
    info = extract_info(s)
    if info:
        return info[0], info[1]
    return "", ""

# This function was previously an AIFunc, but due to its high reusability, it has been staticized as a regular Python function
# This process also demonstrates the potential for AIFunc to be staticized as a regular Python function
def get_swe_debug_task_ctx(task_json_path: str, workspace_path: str) -> SWEDebugTaskCtx:
    ...

    return task_ctx

class SWETaskAIFuncResult(AIFuncResult):
    """
    news result
    """
    debug_task_ctx: SWEDebugTaskCtx = Field(..., description="the detailed information of the swe debug task")

class SWETaskAIFunc(AIFunc):
    """
    get detailed information about the swe debug task
    """
    instruction: str = Field(description="the instruction of the task, to get the detailed information of the swe debug task")

# <moss>
def __aifunc_instruction__(fn: SWETaskAIFunc) -> str:
    return fn.instruction

example = SWETaskAIFunc(instruction="Fetch the metadata of detailed swe debug task information")
# </moss>
\end{lstlisting}

\subsection{ExplorationProjectAIFunc}
ExplorationProjectAIFunc is designed for repository exploration, formulating specific exploration strategies, and returning the code/configuration segments related to the identified issues.

\begin{lstlisting}
# other imports
from ghostos.core.moss import Moss as Parent
from ghostos.core.aifunc.interfaces import AIFunc, AIFuncResult, AIFuncCtx
from evaluation.swe_bench_lite.ai_funcs.file_explorer import FileExplorerAIFunc

@cls_source_code()
class ExplorationProjectAIFuncResult(AIFuncResult):
    found_culprits: Optional[Set[CulpritFilePart]] = Field(..., description="The final culprit file parts, it can be empty if not found")
    cost_steps: int = Field(..., description="The cost steps to find the culprit files")
    confidence_percentage_requirement: int = Field(..., description="The requirement of least confidence percentage of the culprit file part")

class ExplorationProjectAIFunc(AIFunc):
    max_steps: int = Field(default=20, description="the expectation max steps to localize the issue files")
    cur_step: int = Field(default=0, description="the current step of the exploration")
    thoughts: str = Field(default="", description="the brief plan of the exploration")
    debug_task_ctx: Any = Field(..., description="the debug task context")

class Moss(Parent):
    ai_func_ctx: AIFuncCtx
    """useful to run AIFunc"""

# <moss>
def __aifunc_instruction__(fn: ExplorationProjectAIFunc) -> str:
    return (f"ExplorationProjectAIFunc input vals: max_steps: {fn.max_steps}, cur_step: {fn.cur_step}, "
            f"thoughts: {fn.thoughts}, debug_task_ctx: {fn.debug_task_ctx}. You should return an ExplorationProjectAIFuncResult object"
            f"Before you return the culprit file parts, you should use the read_file method or FileExplorerAIFunc, you can also using multi-run or MCTS to explore the key directory. ")
# </moss>    
\end{lstlisting}

\subsection{FileExplorerAIFunc}
This AIFunc is designed for exploration and exploitation operations on a specific file. It was originally defined and created by the ExplorationProjectAIFunc, but due to its high reusability, it has been transformed into a standalone AIFunc. This demonstrates that each layer of AIFunc can create new AIFuncs, allowing the agent to have autonomous extensibility at the intelligent unit level.

\begin{lstlisting}
# imports statements

@cls_source_code()
class FileExplorerAIFuncResult(AIFuncResult):
    found_culprits: Optional[Set[CulpritFilePart]] = Field(..., description="The final culprit file parts, it can be empty if not found")
    confidence_percentage: Optional[int] = Field(..., description="The confidence percentage of the found_culprits is the true culprit")
    file_outline: Optional[str] = Field(default="", description="the important key information or clue, or summarize of the file")

class FileExplorerAIFunc(AIFunc):
    cur_object: str = Field(default="", description="current object to explore")
    debug_task_ctx: Any = Field(..., description="the debug task context")
    file_content: str = Field(default="", description="the content of the file")

class Moss(Parent):
    ai_func_ctx: AIFuncCtx
    """useful to run AIFunc"""

# <moss>
def __aifunc_instruction__(fn: FileExplorerAIFunc) -> str:
    return (f"Your current task is {fn.cur_object}, you should use the read_file method at first, culprit parts might exist in this then thought step by step to find clues about the issue. content of file: {fn.file_content}")
# </moss>    
\end{lstlisting}

\subsection{DebugLocalizationAIFunc Execution}
The following is the main function generated after the lifecycle of MOSS execution:

\begin{lstlisting}
def main(moss: Moss, fn: DebugAgentFn):
    task_json_path = "/home/llm/Project/PythonProjects/GhostOS/evaluation/swe_bench_lite/django_15347.json"
    workspace_path = "/home/llm/Project/PythonProjects/workspace/django"

    # Step 1: Get detailed information about the SWE debug task
    task_fn = SWETaskAIFunc(
        instruction="Get detailed information about the SWE debug task",
        task_json_path=task_json_path,
        workspace_path=workspace_path
    )
    task_result = moss.ai_func_ctx.run("task_info", task_fn)
    
    if not task_result:
        print("Failed to get the task information.")
        return None, False
    
    debug_task_ctx = task_result.debug_task_ctx

    # Step 2: Prepare the repository for debugging
    repo_manager = RepositoryContextManager(debug_task_ctx)
    prepare_result = repo_manager.prepare_repository_for_debug()

    if prepare_result != PrepareRepositoryResult.PREPARE_SUCCESS:
        print("Failed to prepare the repository for debugging: ", prepare_result)
        return None, False

    # Step 3: Localize the issue files using ExplorationProjectAIFunc
    exploration_fn = ExplorationProjectAIFunc(
        max_steps=20,
        cur_step=0,
        thoughts="Starting localization of the issue files",
        debug_task_ctx=debug_task_ctx
    )
    exploration_result = moss.ai_func_ctx.run("exploration", exploration_fn)
    
    if not exploration_result:
        print("Failed to localize the issue files.")
        return None, False
    
    issue_culprit_file_parts = exploration_result.found_culprits

    # Step 4: Reset the repository after debugging
    repo_manager.reset_repository_after_debug()

    # Create the result
    result = DebugAgentFnResult(
        issue_culprit_file_parts=issue_culprit_file_parts,
        err=None
    )
    
    return result, True    
\end{lstlisting}
This is an overarching outline of the thought process. We can see that the LLM has planned four steps, each implemented through an AIFunc. Each AIFunc represents a conceptual thought, defining input parameters and return values, while the connections between them are realized through Python code, reflecting the thought process. Since the inputs and outputs of the AIFuncs are typed, type checking can be conveniently performed. The intermediate Python code specifically embodies the thought and its implementation. This design achieves Turing completeness in scenarios involving multiple rounds of thought and execution through synchronous calls.

When the DebugLocalizationAIFunc reaches the line `task\_result = moss.ai\_func\_ctx.run(key="swe\_task", fn=swe\_task\_fn)`, it dives into the writing and execution process of the next AIFunc.

\subsection{SWETaskAIFunc Execution}

\begin{lstlisting}
def main(moss: Moss, fn: SWETaskAIFunc):
    try:
        # Retrieve the task context using the provided function
        task_ctx = get_swe_debug_task_ctx(fn.task_json_path, fn.workspace_path)
        
        # Prepare the result
        result = SWETaskAIFuncResult(debug_task_ctx=task_ctx)
        return result, True
    except Exception as e:
        logging.error(f"Error occurred: {e}")
        return None, False    
\end{lstlisting}
Upon completion of the execution, return the SWETaskAIFuncResult to the upper-level DebugLocalizationAIFunc and terminate the current AIFunc execution.

\subsection{ExplorationProjectAIFunc Execution}
This is the AIFunc responsible for exploring the repository. It employs various search strategies, including static pure Python code and dynamic AIFuncs, to identify the root cause files related to the issue.

The generated main function is as follows:

\begin{lstlisting}
def main(moss: Moss, fn: ExplorationProjectAIFunc):
    # Define useful constants
    issue_info = fn.debug_task_ctx.issue_info
    workspace_path = fn.debug_task_ctx.workspace_path
    repo = fn.debug_task_ctx.repo
    
    # Define the directory to start exploration
    start_directory = f"{workspace_path}/{repo}"
    
    # Initialize the FileContentOperations
    file_operations = FileContentOperations(workspace_path)

    # Define the search string from the issue description
    search_string = "extra_tags"
    
    # Find files containing the search string in the repository
    files_containing_string = file_operations.find_files_containing_string_in_directory(start_directory, search_string)
    
    if not files_containing_string:
        print("No files found containing the search string.")
        return None, False

    # Initialize variables for exploration
    found_culprits = set()
    cur_step = fn.cur_step
    max_steps = fn.max_steps
    confidence_percentage_requirement = 70  # This is an assumed requirement

    # Explore each file containing the search string
    for file_path in files_containing_string:
        if cur_step >= max_steps:
            break

        # Read the content of the file
        file_content = file_operations.read_file(file_path)
        
        # Create an instance of FileExplorerAIFunc to analyze the file
        file_explorer = FileExplorerAIFunc(
            cur_object=file_path,
            debug_task_ctx=fn.debug_task_ctx,
            file_content=file_content
        )
        
        # Execute the FileExplorerAIFunc
        file_explorer_result = moss.ai_func_ctx.run(f"explore_{file_path}", file_explorer)
        
        if isinstance(file_explorer_result, FileExplorerAIFuncResult):
            if file_explorer_result.confidence_percentage >= confidence_percentage_requirement:
                for culprit in file_explorer_result.found_culprits:
                    found_culprits.add(culprit)
        
        cur_step  = 1

    # Create the result object
    result = ExplorationProjectAIFuncResult(
        found_culprits=found_culprits,
        cost_steps=cur_step,
        confidence_percentage_requirement=confidence_percentage_requirement
    )

    return result, True    
\end{lstlisting}

The search strategy implemented by this AIFunc focuses on traversing all files that contain the string `extra\_tags`, as specified in the issue description. It employs the FileExplorerAIFunc to conduct a detailed analysis of each file. A file is deemed a root cause of the issue if it achieves a confidence percentage exceeding 70. This approach exemplifies the dynamic nature and flexibility of the MOSS + AIFunc integration.

When the command moss.ai\_func\_ctx.run(f"explore\_\{file\_path\}", file\_explorer) is executed, it initiates the process of editing and executing the FileExplorerAIFunc.  

\subsection{FileExplorerAIFunc Execution}
The execution process of each FileExplorerAIFunc is quite similar; here is an example of one. First, it reads the content of the file, then analyzes the issue description and returns a FileExplorerAIFuncResult that includes the found culprits and the confidence percentage.

\begin{lstlisting}
def main(moss: Moss, fn: FileExplorerAIFunc):
    # Initialize FileContentOperations with the workspace path
    file_ops = FileContentOperations(fn.debug_task_ctx.workspace_path)
    # Read the content of the file, starting with page 1
    file_content = file_ops.read_file(fn.cur_object, page_number=1)
    # Update the file_content attribute of the FileExplorerAIFunc instance
    fn.file_content = file_content
    # Return the updated FileExplorerAIFunc instance and indicate the operation was successful
    return fn, True    
\end{lstlisting}

\subsection{Task Result}
After reviewing several files, the final LLM returns the root cause file, restores the environment, and outputs the results to the console:
\begin{lstlisting}
Previous HEAD position was 7c4f396509 Stopped including type="text/css" attributes for CSS link tags.
Switched to branch 'main'
Your branch is up to date with 'origin/main'.
Deleted branch django__django-15347 (was 01a4d8a3c7).
AgentFnResult(issue_files=['/home/llm/workspace/django/django/contrib/messages/storage/cookie.py'], err=None)    
\end{lstlisting}

Although we have not yet conducted extensive testing of the debugging effectiveness on benchmarks like SWEBench, the exploration process described above demonstrates that the combination of MOSS and AIFunc can conveniently create a software debugging agent. The intelligent components exhibit excellent flexibility and scalability, giving the impression that the entire workflow has already been established, when in fact, we have only defined a few initial AIFuncs.

\end{document}